\documentstyle[prb,aps,eqsecnum]{revtex}
\begin{document}
\title{Spontaneous Currents in Josephson Devices}
\author{Z. Radovi\'c,$^1$ L. Dobrosavljevi\'{c}--Gruji\'{c},$^2$ B. Vuji\v{c}i\'c$^3$}
\address{$^1$ Faculty of Physics, University of Belgrade, P.O. Box 368, 11001 Belgrade, Yugoslavia}
\address{$^2$ Institute of Physics, P.O. Box 57, 11001 Belgrade, Yugoslavia}
\address{$^3$ Faculty of Science, University of Montenegro, P.O. Box 211, 81000 Podgorica, Yugoslavia}
\address{\bf To be published in Phys. Rev. B {\bf 60} (1 september 1999-I)}

\maketitle

\begin{abstract}
The  unconventional  Josephson   coupling in a ferromagnetic weak link between
$d$-wave superconductors  is
studied theoretically.
 For strong ferromagnetic barrier
influence, the unconventional coupling, with ground state phase
difference across the link $0<\phi_{\rm gs}\leq \pi$, is obtained  at
small crystal misorientation of the superconducting electrodes,
 in contrast to the case of normal metal
barrier, where it appears at large misorientations.
In both cases, with decreasing temperature there is an increasing range of
misorientations,  where $\phi_{\rm gs}$
 varies continuously between $0$ and $\pi$.
   When the weak
link is a part of a superconducting ring, this is accompanied by
the flow of spontaneous supercurrent, of intensity which depends (for
a  given misorientation) on the reduced inductance $l=2\pi LI_c(T)/\Phi_0$,
and is non-zero only for $l$ greater than a critical value. For $l\gg
1$, another consequence of the unconventional coupling is the
anomalous  quantization of the magnetic flux.
\end{abstract}

\pacs{PACS numbers: 74.50+r, 74.80Fp}

\section{Introduction}

Magnetic scattering effects and $d$-wave pairing in superconductors
may have similar manifestations, such as $\pi$-phase states
and spontaneous currents in  Josephson devices.\cite{spie}

The phase  of the superconducting order parameter as a function of
the momentum-space direction is different for $d$-wave and $s$-wave
pairing states, since the latter has a single phase, whereas the $d$-wave
state exhibits jumps of $\pi$ at the (110) lines.\cite{grenoble} The
Josephson current through a junction depends on the phase difference
$\phi$ between the superconductors on the either side, $I \sim \sin \phi$,
and the negative sign to the prefactor  of $I$ can be regarded as an
additional phase shift of $\pi$. Superconducting rings containing an
odd number of such $\pi$ shifts have the spontaneous magnetization
$\pm \Phi_0/2$, and quantized flux of $(n+1/2)\Phi_0$, where $n$ is
an integer, providing that the product of the critical current and
the self-inductance of the ring is $I_c L \gg \Phi_0/2\pi$.\cite{t}

Beside $d$-wave pairing, two alternative mechanisms for
$\pi$-phase shifts at a conventional junction have been proposed.
One is the spin-flip scattering in the junction  barrier containing
paramagnetic impurities,\cite{5t} and the other is the indirect
tunneling through a localized state, in which correlation effects
produce a negative Josephson coupling between two superconducting
grains.\cite{6t}

A  problem related to the first mechanism is the possibility of
$\pi$-coupling in superconductor/ferromagnetic metal (S/F) weak links
and multilayers, due to the presence of the exchange field in
F.\cite{buzdin,zoran,vesna,demler} The characteristic oscillations of the critical
temperature $T_c$ of multilayers  as a function of the ferromagnetic
 layer thickness
$d_{\rm F}$,  predicted theoretically by Radovi\'c et al.
\cite{zoran} were observed  in Nb/Gd multilayers by Strunk et
al.\cite {strunk} and by Jiang et al.\cite {drag} and in Nb/CuMn
(superconductor/spinglass multilayers) by Mercaldo et al.\cite {mer}
While the oscillatory $T_c$ behavior was interpreted in terms of
$\pi$-coupling in Refs. \onlinecite{drag} and \onlinecite{mer}, in
Ref. \onlinecite{strunk} it was attributed to the change from
paramagnetic to ferromagnetic state with increasing $d_{\rm F}.$
 Recently,
double minimum $T_c$ oscillations were observed on Nb/Co and V/Co
multilayers by Obi et al.,\cite{obi} and the second minimum was found
to be consistent with the appearance of the $\pi$-phase.

In connection with the second mechanism,  it was predicted that the
anomalous flux quantization may appear with $50 \% $ probability
in rings made from either disordered superconductors, or granular
superconductors doped with paramagnetic impurities.  The latter
possibility was tested recently on uniform Mo rings doped with Fe,
but the absence of the effect of paramagnetic impurities on flux
quantization was found.\cite{tinkham} This may be due to the fact
that the phase shifts induced by the impurities do not survive
averaging over different electron paths around the ring. It was
suggested that a loop with nanometer size point contact, doped with
magnetic impurities, might satisfy the requirement that electrons
circling the loop are subject to the same exchange potential.

The Josephson effect in weak links with ferromagnetic metal barrier
can be the direct probe of unconventional coupling.  Tunneling properties of $s$-wave
junctions with ferromagnetic insulator barrier have been studied theoretically by
Kuplevakhskii and Fal'ko,\cite{ku-t} De Weert and Arnold,\cite{arnold}
and more recently, by Tanaka and Kashiwaya.\cite{tanaka-c} In the
latter work the possibility of $\pi-$ shift was obtained.  Tanaka and
Kashiwaya studied separately the Josephson effect in  anisotropic
superconductor junctions with any symmetries and with a nonmagnetic
insulating barrier.\cite{tanaka} This reference provides a general
result from which several existing theories can be derived, and
contains an extensive list of relevant references. Short weak links
and  $d$-wave junctions
with normal metal barrier  were studied by Barash
et al.\cite{barash1,barash2} These authors find an essentially
nonharmonic current-phase relation $I(\phi)$ at low temperature, and
a flow of spontaneous
supercurrents in superconducting ring
interrupted by $\pi-$ junctions.  Recently, Fogelstrom et al. have
considered  pinhole junctions in $d$-wave superconductors and have
shown that at low temperature the ground state phase difference at
the junction may vary continuously between $0$ and $\pi$.
\cite{fogelstrom}

In the present work, we show how new possibilities for unconventional
coupling in S/F/S Josephson devices result from the combined  effects
of exchange field and $d$-wave pairing. We consider a Josephson weak
link in the clean limit, with thin and short ferromagnetic (or
normal) metal barrier. The interfaces between the barrier and the
superconducting electrodes are assumed fully transparent.
Two-dimensional (2D) superconductivity and $d_{x^2-y^2}$ symmetry of the
order parameter are considered.\cite{maki} Assuming  the barrier
 perpendicular to the $a-b$ planes of
superconducting electrodes,  the influence of their relative
orientation  is studied. For comparison some
results for $3D$ $s$-wave case are included.
  In Section II, we give a brief overview of
the quasiclassical theory of superconductivity in presence of
exchange field and for anisotropic pairing interaction, which we apply
in the following.  In Section III, we calculate the quasiclassical
Green's functions for the weak link to obtain the supercurrent as a
function of the phase difference at  the contact.  When the weak link
is a part of a superconducting ring, the ground state phase
difference is calculated and the conditions for the flow of
spontaneous supercurrent and anomalous flux quantization are
obtained. Section IV contains a discussion of the results and a brief
conclusion.

\section{Quasiclassical Equations}

A microscopic approach powerful enough  to deal with
superconductivity in  restricted geometries is the Eilenberger
quasiclassical theory of superconductivity. \cite{alexander} To
calculate the critical current of $d$-wave  S/F/S contact, we
generalize the method developed for the su\-per\-con\-duc\-tor/normal
metal/su\-per\-con\-duc\-tor (S/N/S) contact by Svidzinskii and
Likharev,\cite{svi,lih} applied to  S/F/S contact for $s$-wave pairing
case by Buzdin et al.\cite{buzdin} and by Demler et al.\cite{demler}

The basic equations of the Eilenberger quasiclassical  theory
of superconductivity, in the presence of an exchange energy $h$, are
\begin{equation}
[i\omega_n\hat\tau_3-{\bf {\hat \Delta}}+h\hat\sigma_z \hat 1,\, \hat g]+i
{\bf v}_0\cdot\nabla\hat g = 0\;,
\label{matrixeq}
\end{equation}
with normalization condition,
\begin{equation}
\hat g^2 =- \pi^2\hat 1\;,
\label{norm}
\end{equation}
where
$\omega_n = \pi T(2n+1)$ are the Matsubara frequencies ($\hbar =k_B=1$),
${\bf v}_0$ is the Fermi velocity, and
$\hat g = \hat g({\bf v}_0,{\bf R};\omega_n)$ is the quasiclassical Gorkov's Green's
function integrated over energy.
Explicitly,
\begin{equation}
\hat{g} = \pi\left[
\begin{array}{cccc}
-ig_\uparrow&0&0& f_{\uparrow\downarrow}\\
0&-ig_\downarrow&- f_{\downarrow\uparrow}&0\\
0& f_{\uparrow\downarrow}^+&ig_\downarrow&0\\
- f_{\downarrow\uparrow}^+&0&0&ig_\uparrow
\end{array}
\right],
\label{matrixg}
\end{equation}
the gap matrix is
\begin{equation}
{\bf \hat \Delta }=
\left[\begin{array}{cc}
0&i\sigma_y\Delta({\bf v}_0,{\bf R})
\\[4pt]
i\sigma_y\Delta^\ast(-{\bf v}_0,{\bf R})
&0
\end{array}
\right],
\label{matrixdelta}
\end{equation}
and the self-consistency equation is
\begin{equation}
\Delta({\bf v}_0,{\bf R}) = \pi N T\sum_{\omega_n}
\left\langle V({\bf v}_0,{\bf v}_0')
{ f_{\downarrow\uparrow}({\bf v}_0',{\bf R};\omega_n) +
f_{\uparrow\downarrow}({\bf v}_0',{\bf R};\omega_n)  \over 2 }
\right\rangle,
\label{(2.5)}
\label{selfconst}
\end{equation}
where $V({\bf v}_0,{\bf v}_0')$ is the pairing interaction.
The supercurrent density is given by
\begin{equation}
{\bf j}({\bf R}) = -2ie\pi N T\sum_{\omega_n}
\left\langle {\bf v}_0 {g_\uparrow
+g_\downarrow\over2}\right\rangle ,
\label{currenteq}
\end{equation}
where $\langle\cdots \rangle$  denotes the angular averaging over
the Fermi surface, $N$ being the density of states at the Fermi surface.

In the following, we use the notation $ f_{\downarrow\uparrow} = f$,
$f_{\uparrow\downarrow}^+ = f^+,$ $g_{\downarrow}=g $ for the set of
Green's function for one spin direction (down with respect  to the
exchange field orientation).  They satisfy the   scalar equations
\begin{equation}
2(w_n + ih)f+ {\bf v_0}\nabla f=2g\Delta \label{(2.7)}, \\
\end{equation}
\begin{equation}
2(w_n + ih)f^+ -{\bf  v_0}\nabla f^+ =2g\Delta^* \label{(2.8)}, \\
\end{equation}
\begin{equation}
{\bf v_0}\nabla g = \Delta^* f - f^+\Delta.    \label{(2.9)}
\end{equation}

For the opposite spin direction, the corresponding set of Green's
functions is obtained by changing $h\to -h$.

\section{Solutions}

We solve  the quasiclassical equations for a $d$-wave S/F/S junction, where
S is  an anisotropic superconductor with $d_{x^2-y^2}$ symmetry,
and F is a monodomain ferromagnetic metal with constant exchange
energy $h$.

We assume  both S and F metals clean, with same dispersion
relations and with same Fermi velocity ${\bf v_0}$. Electron scattering
on impurities in S  can be neglected if  $l\gg \xi_0$,
and in F if $h\gg  v _0/l$,
 where $l$ is the
electron mean free path and $\xi _0$
 the BCS superconducting coherence length.

The thin and short barrier, of thickness $d$, is assumed
perpendicular to the $\hat a$ axis in the $\hat a-\hat b$ plane of
the lhs monocrystal $\rm S_L$, which may be misoriented with respect
to the rhs one, $\rm S_R$, their $\hat a$ axes making an angle
$\theta$ (Fig. 1). For anisotropic pairing, the pair potential and
the shape of quasiparticle spectra depend on the misorientation.  For
$d_{x^2-y^2}$ symmetry, \cite{maki} the pairing interaction and the
pair potential in $\rm S_L$ are $V({\bf {v_0}, \bf {v_0'}}) \propto
\cos 2\varphi \cos 2\varphi'$ and $\Delta (\bf {v_0})\propto \cos
2\varphi$, respectively,  where  $\varphi$ is the angle the
quasiparticle momentum makes with the $\hat a$ axis.  Similarly, in
$\rm S_R$, $\Delta ({\bf {v_0}})\propto \cos 2(\varphi - \theta).$
Assuming  2D nature of  HTS with $d$-wave pairing, we take
cylindrical Fermi surface for both metals, whereas for 3D $s$-wave
case spherical Fermi surface is taken.

To solve Eqs.~(\ref{(2.7)})--(\ref{(2.9)}),
we take in F, where $h=const.$,  $\Delta =0$, but $f\neq 0$ due to
proximity of S. In S,
where $h=0$, we take the order parameter on two sides of the  barrier
in the form $\Delta_{\rm L,R}=\Delta_{\rm L,R} (\varphi ) e^{\pm i \phi/2}$,
where $\phi $
is the phase difference at the contact,
$\Delta_{\rm L}(\varphi )=\Delta_0\cos 2\varphi$,
$\Delta_{\rm R}(\varphi )=\Delta_0\cos 2(\varphi -\theta)$
for $d$-wave ($\Delta (\varphi)=\Delta _0$  for
$s$-wave) pairing.\cite{zika}
Assuming $d/\xi_0<1$, we neglect the spatial variation of $\Delta$,
which in principle could be obtained from the self-consistency equation
(\ref{(2.5)}).

Far from the barrier, $f$ and $g$ approach their
respective bulk values
\begin{equation}
\langle f_{\rm L,R}\rangle  = \langle f_{\rm L,R}^+\rangle ^\ast =
 {\Delta_{\rm L,R}/\tilde \omega_{n \rm L,R} },
\end{equation}
and
\begin{equation}
\langle g_{\rm L,R}\rangle  = {
\omega_n /\tilde \omega_{n \rm L,R} }.
\end{equation}
where
\begin{equation}
  \tilde \omega_{n \rm L,R}
= \sqrt{\omega_n ^2+|\Delta|_{\rm L,R}^2}.
\end{equation}
Choosing the $x$-direction along ${\hat a}_{\rm L}$,
we look for the solution at the rhs, $ x\geq d/2$,
\begin{equation}
 f_{\rm R} = \langle f_{\rm R}\rangle +\bar f_{\rm R}e^{-\alpha_{\rm R} x},
\quad g_{\rm R} = \langle g_{\rm R}\rangle +\bar g_{\rm R}e^{-\alpha_{\rm R} x}{.}
\label{rhs}
\end{equation}
Similarly, at lhs, $x\leq d/2$,
\begin{equation}
 f_{\rm L} = \langle f_{\rm L}\rangle +\bar f_{\rm L}e^{\alpha_{\rm L} x},
\quad g_{\rm L} = \langle g_{\rm L}\rangle +\bar g_{\rm L}e^{\alpha_{\rm L} x}.
\label{lhs}
\end{equation}
From Eqs. (\ref{(2.7)})--(\ref{(2.9)}) we find
\begin{equation}
\bar f_{\rm R} = \bar g_{\rm R}{\Delta_{\rm R}\over \omega_n -\alpha_{\rm R} {\rm v}_x/2},\quad
\bar f_{\rm L} = \bar g_{\rm L}{\Delta_{\rm L}\over \omega_n +\alpha_{\rm L}
 {\rm v}_x/2}\;,
 \label{frfl}
\end{equation}
with
\begin{equation}\alpha_{\rm L,R} =
 {2\tilde \omega_{n \rm L,R} \over {\rm v}_x} =
  {2\tilde \omega_{n \rm L,R} \over {\rm v}_0\cos\varphi}\;.
\label{alpha}
\end{equation}

For F,  $ \Delta = 0$, and the solutions are
\arraycolsep=1pt
\begin{equation}
\begin{array}{rl}
f &=\displaystyle  C\exp\left\{\frac{-2(\omega_n +ih)x}{\rm v_{x}}\right\},\\[12pt]
f^+ &=\displaystyle  C^+\exp\left\{\frac{2(\omega_n +ih)x}{\rm v_{x}}\right\},
\end{array}
\label{ferrosolf}
\end{equation}
and
$g = \hbox{const.}$ for $|x|\leq d/2$. Note that
$f^+ = f(-{\bf v}_0,\Delta^\ast).$
Assuming a transparent boundary between two metals,
we use  the continuity conditions for $g$ and $f$ at the
barrier interfaces, $x = \pm d/2$.
We find the normal Green's function in the barrier
\widetext
\begin{equation}
g=
\frac{\omega_n}{\tilde{\omega_n}_{\rm L}}
-{{(\omega_n+\tilde \omega_{n{\rm L}})[(\omega_n-\tilde \omega_{n{\rm R}}){\Delta_{\rm L}}(\varphi)
-(\omega_n-\tilde \omega_{n{\rm L}}){\Delta_{\rm R}}(\varphi)\exp( i\chi )]}\over
{\tilde \omega_{n{\rm L}}[
 (\omega_n-\tilde \omega_{n{\rm R}}){\Delta_{\rm L}}(\varphi)
-(\omega_n+\tilde \omega_{n{\rm L}}){\Delta_{\rm R}}(\varphi)\exp( i\chi )]}},
\label{gnb}
\end{equation}
%
where
\begin{equation}
\chi = \phi+\frac{Z}{\cos\varphi} - \frac{2i\omega_n d}{{\rm v}_0\cos\varphi}\;
\label{kappa}
\end{equation}
and
\begin{equation}
Z=\frac{2hd}{{\rm v}_0}
\label{Z}
\end{equation}
is the parameter measuring the ferromagnetic barrier influence.
The temperature dependence of $\Delta _0=\Delta _0 (T)$ is quite similar
 to the BCS one, and can be approximated by
\begin{equation}
\Delta _0 (T)=\Delta _0 (0)\tanh\left(1.74\sqrt {T_c/T -1}\right).
\label{deltat}
\end{equation}
The only difference  is that in the case of $d$-wave pairing
$\Delta _0 (0)/\Delta _0^{BCS}(0)>1$. \cite{maki}

From the above results it is easy to obtain the corresponding ones
for $d$-wave junction with normal metal barrier, putting $h\to 0$, and
for $s$-wave junction, taking isotropic pair potential
$\Delta^{\rm L}=\Delta^{\rm R}=\Delta_0$, with ferromagnetic ($h\neq 0$) or
normal metal ($h=0$) barrier.

\subsection{Supercurrents}

In S/N/S case, for $d$-wave pairing the ground state phase difference
on the contact is zero, as for the $s$-wave pairing, if two  S
monocrystals have the same orientation with respect to the barrier.
This may not be the case for S/F/S contacts, and we calculate the
supercurrent as a function of the barrier exchange field intensity,
as well as the function of the orientation.

Taking the current direction along $\hat a_L$, from Eq.~(\ref{currenteq})
 we get for 2D $d$-wave case
\begin{equation}
I = I_0 t\sum_{\omega_n,\sigma} \int^\infty_1
\frac{\Im [g_{\sigma}(u)
+g_{\sigma}(-u)]}{u^2\sqrt{u^2-1}}\,{\rm d}u,
\label{jxgen}
\end{equation}
where $u=1/\cos\varphi$ and $\sigma=\downarrow, \uparrow$.
Green's function $g_{\downarrow}$ is given by Eq. (\ref{gnb}),
and $g_{\uparrow}(h)=g_{\downarrow}(-h)$.
Eq.~(\ref{jxgen}) gives the
supercurrent through the barrier of the area $S$, $I=jS$
as a function of $\phi$, $\theta$, temperature $T$
{\it via}  $t=T/ \Delta _0 (T)$, and
of parameters measuring the influence of the F barrier $Z$
 and $\tilde d = d/\xi_0(0)$, where $\xi_0(0)=
{\rm v}_0/\pi \Delta_0(0)$. Note that $Z\sim {\tilde d} h /\Delta_0(0)$.
The  temperature dependent normalizing current is
\begin{equation}
I_0=\frac{2\Delta_0(T)}
{e R_{\rm N}},
\label{I0}
\end{equation}
where the normal resistance is given by
$R_{\rm N}^{-1}=e^2{\rm v}_0NS$,
$N$ being the density of states at the Fermi surface.

In the limit  $T\to T_c$, numerical calculations show that
Eq.~(\ref{jxgen}) gives, as expected,
the Josephson relation
\begin{equation}
I=I_c \sin \phi,
\label{jxsin}
\end{equation}
where the sign and magnitude of $I_c$  depend on $T$,
$\theta$ and $Z$.

For 3D case and $s$-wave pairing,\cite {buzdin}  Eq.~(\ref{gnb}) reduces to
\begin{equation}
g={{\omega_n\cos{\chi \over 2}+i\tilde \omega_n\sin{\chi \over 2}}\over
{\tilde \omega_n\cos{\chi \over 2}+i \omega_n\sin{\chi \over 2}}}\,,
\label{3d}
\end{equation}
where
\begin{equation}
\tilde \omega_n=\sqrt{\omega_n^2+|\Delta_0|^2}\,,
\end{equation}
and Eq.~(\ref{currenteq}) gives
\begin{equation}
I = \frac{\pi}{4}\,I_0 t \sum_{\omega_n,\sigma}
\int^\infty_1 {\Im [g_{\sigma}(u)
+g_{\sigma}(-u)]\over
u^3 }\,{\rm d}u,
\label{jx3D}
\end{equation}
where the same notation as in Eq.~(\ref{jxgen}) is used.

In general, the supercurrents in the Josephson junctions and
 weak links are carried by
the Andreev bound states in the barrier.\cite{vesna,tanaka,furu}
In the present case,
 the spectrum
of these states    has been calculated from the analytical continuation of
the above Green's functions $g_{\sigma}$.\cite{zikic}

\subsection{Magnetic flux}

For a superconducting ring containing one Josephson junction, the
ground state phase difference
$\phi_{\rm gs}$  at the junction
can be obtained by minimizing the reduced energy \cite{barone}
\begin{equation}
\tilde W(\phi)=\frac{1}{I_c}\int_0^\phi  I(\phi')\,{\rm d}\phi'
+ \frac{1}{2}l^{-1}(\phi-\phi_{\rm e})^2,
\label{energy}
\end{equation}
where  $I(\phi)$ is calculated from Eq.~(\ref{jxgen}),
or Eq.~(\ref{jx3D}), $\tilde{W}=W/(\Phi_0I_c/2\pi)$,
$\Phi_0$ is the flux quantum, and the reduced inductance
is
\begin{equation}
l= \frac{2\pi}{\Phi_0}\,I_c L
\label{induct}
\end{equation}
with $I_c=I_c(T,\theta, Z)$.
For $l^{-1}=0$, the ground state condition reduces to
${I}(\phi_{\rm gs})=0$,
${I}{}'(\phi_{\rm gs})>0$.

In the limit $T\to T_c$, where the current-phase relation is harmonic,
Eq.~(\ref{jxsin}), from Eq.~(\ref{energy}) it follows
\begin{equation}
\sin \phi_{\rm gs} + l^{-1}(\phi_{\rm gs}-\phi_{\rm e})=0.
\label{harm}
\end{equation}

The phase $\phi_{\rm e}=2\pi \Phi_{\rm e}/\Phi_0$ represents the reduced external
magnetic
flux through the ring. In the ground state, the total magnetic flux
\begin{equation}
\Phi={\Phi_0\over 2\pi}\,\phi_{\rm gs}
\label{flux}
\end{equation}
is related to the spontaneous supercurrent
$I_{\rm gs}=I(\phi_{\rm gs})$ by
\begin{equation}
\Phi=\Phi_{\rm e} - {\Phi_0 \over 2 \pi}\,l I_{\rm gs}.
\label{total}
\end{equation}
Note that $\phi_{\rm gs}$ now depends on $\phi_{\rm e}$ and
when\break  $n\leq {\Phi/\Phi_0}\leq {n+1}$,
$n=1,2,3\dots$, $\phi_{\rm gs}\to \phi_{\rm gs} +2\pi n$.

\section{Results and conclusion}

To illustrate the consequences of unconventional coupling we present
the results of numerical calculations  for  typical
cases, low temperature, $t=0.05$ ($T/T_c\approx 0.1$) and  high
temperature, $t=5$ ($T/T_c\approx 0.9$), for  strong influence of the
ferromagnetic barrier ($Z=3$) and for  the normal metal barrier
($Z=0$).

For $d$-wave pairing, $I(\phi)$ dependence is much more complex than
in $s$-wave pairing case, even in the absence of exchange field.
Besides deformations of sinusoidal curves at low $T$, the shape and
sign of $I(\phi)$ depend very much on the misorientation angle
$\theta$, both for normal  and ferromagnetic metal  barriers.  The low $T$ deformations of $I(\phi)$ are reminiscence
of discontinuities at $T=0$.  At zero temperature,   for $Z=0$ the
current turns out to be discontinuous at $\phi=\pi$ for $\theta=0$,
as in the $s$-wave pairing case.\cite{lih} For $\theta=\pi/2$, the
current jump is at $\phi=0$ (or $2\pi$), and  for $\theta=\pi/4$
smaller jumps appear both at $\phi=0$ and $\pi$, similarly to the
results of Ref.  \onlinecite{barash1}.  With
increasing temperature, $I(\phi)$ becomes less deformed tending to
sinusoidal variations. Similar conclusions hold for
$Z=3$. The dependence $I(\theta)$ persists in the
presence of exchange field, as a strong evidence of $d$-wave pairing.
This is shown in Fig. 2 for the characteristic values of $Z$,  $t$
and  $\theta$.

In the absence of external magnetic field, for  $\theta=0$ the
ground state phase difference $\phi_{\rm gs}$ at the S/F/S weak link
is always zero or $\pi$, both for 3D $s$-wave  and for
2D $d$-wave pairing.  For weak influence of the ferromagnetic
barrier, $Z\lesssim 1$, $\phi_{\rm gs}=0$ and for stronger influence,
$1\lesssim Z\lesssim 4$, there is a $\pi-$shift, $\phi_{\rm gs}=\pi$ (Fig.3). Larger
values of $Z$, for which $\phi_{\rm gs}=0$ again, would correspond to
the decoupling of S electrodes.\cite{ku-d} For $\theta=\pi/2$,
 $\phi_{\rm gs}=0 \to \phi_{\rm gs}=\pi$  and {\it vice versa}. These results are temperature independent.
  For $s$-wave tunnel
junctions with ferromagnetic insulating barrier it is also found
that $\phi_{\rm gs}$ changes from $0$ to $\pi$ as the exchange
interaction is enhanced.\cite{tanaka-c} In weak links, for 2D
$d$-wave pairing   we find that $\phi_{\rm gs}$
varies between $0$ and $\pi$, depending on $\theta$ and $Z$. This
variation is monotonous  at low temperature, whereas at high
temperature we find step function-like jumps at $\theta=\pi/4$.
In the vicinity of $\theta=\pi/4$, depending on $Z$ and  $\theta$  the transition from
some intermediate value, $0<\phi_{\rm gs}<\pi$, to $\phi_{\rm gs}=0$
or to $\phi_{\rm gs}=\pi$ may  occur with the change of temperature,
similar to the $0$ to $\pi$ transition found for $d$-wave tunnel
junction,\cite{barash2} see Figs. 4(a) and 4(c).  For $d$-wave
pinhole junctions a range of crystal orientations, where $\phi_{\rm
gs}$ varies from $0$ to $\pi$, is also found at low
$T$.\cite{fogelstrom}

 For superconducting rings with  sufficiently large
reduced  inductance $l$, interrupted by the S/F/S link, the
unconventional coupling is also found, with nonzero $\phi_{\rm
gs}<\pi$, which rapidly decreases with increasing $l^{-1}$,  Figs.
4(a) and 4(c). We note that at high temperature, $t>0.5$,
 $\phi_{\rm gs}\equiv 0$  for $l^{-1}\geq 1$.

The appearance of spontaneous current $I_{gs}(\theta)$  in superconducting
rings for external magnetic flux $\Phi_{\rm e}=0$  is illustrated in
Figs. 4(b)  and 4(d) for several values of the reduced inductance.
The flow of spontaneous supercurrent is the consequence of the
misorientation effect, or of the exchange field influence, whenever
$\phi_{\rm gs} \neq 0$.  This is a generalization of the spin-flip
induced effect predicted by Bulaevskii et al. for conventional
superconductor rings with  junction  barrier doped with paramagnetic
impurities.\cite{5t} The highest values of spontaneous currents,
corresponding to $\theta=\pi/2$ for $Z=0$, and to $\theta=0$ for
$Z=3$, strongly depend on the reduced inductance.  Maximum values,
equal to the critical currents, correspond to $l^{-1}\sim 1$ at low $T$,
 and to $l^{-1}<1$ at high $T$ (Fig. 5).
 It is important to point out that the spontaneous
currents appear only for sufficiently large  reduced inductance,
greater than some critical value $l_c$, in contrast to the result
of Barash et al. for S/N/S junctions at $T=0$.\cite{barash1} At high $T$,
$t>0.5$, where the current-phase relation is harmonic, we obtain an universal curve
 below $l_c^{-1}=1$, as in the case of tunnel junctions.\cite{5t} At low $T$,
 the spontaneous currents flow is obtained in a wider range of $l^{-1}$,
 the shape of the curves and $l_c$ depending on the barrier influence
   $Z$ and on
 the type of pairing symmetry.

In the presence of external magnetic flux, $\Phi_{\rm e}\neq 0$, the
flow of spontaneous current leads to the anomalous flux quantization
in the ring, both for S/N/S and S/F/S junctions.   In the
first case, $Z=0$, this occurs in the vicinity of $\theta = \pi/2$
and in the second case, $Z=3$, in the vicinity of  $\theta = 0$.  The
effect of half magnetic flux quantization, $\Phi/\Phi_0=1/2,
3/2,\dots$,  pronounced for $l^{-1}$ small, becomes smeared out and
eventually lost for larger $l^{-1}$ (Fig. 6).

  Since
 $l^{-1}\propto 1/I_c(T)$ rapidly
increases with increasing temperature, the unconventional coupling
effects, such as the spontaneous supercurrent flow and the anomalous
flux quantization exist only below some  temperature   characteristic
for the given device.

In conclusion, the spontaneous supercurrent flow in a HTS ring with
ferromagnetic weak link provides new possibilities for experimental
observation of the unconventional Josephson coupling, $0<\phi_{\rm
gs}\leq \pi$.  Due to the presence of the exchange field in the
barrier, it may appear, manifested by the anomalous flux
quantization, without the misorientation of the crystals in two
superconducting electrodes.  However, the exchange field effect do
not mask the difference between $s$ and $d$-wave pairing, since the
coupling depends on the misorientation. We emphasize that the
unconventional coupling effects in the  weak links
 are much more diverse than in the tunnel junctions, which display
in the ground state only $0$ or $\pi$ phase difference.

\newpage

\begin{figure}
\caption{Schematic illustration of the weak link.
The fourfold symmetry of the gap functions is indicated.}
\end{figure}

\begin{figure}
\caption{Reduced current $I/I_0$ as a function of phase difference
$\phi$, for S/N/S, $Z =0$  and
S/F/S, $Z =3$ weak links, at low temperature
$t=0.05$, and at high temperature $t=5$. The reduced
thickness is $\tilde d=0.25$. 2D $d$-wave pairing for three values of the
misorientation angle $\theta$ (solid curves), and 3D $s$-wave pairing
(dashed curves).}
\end{figure}

\begin{figure}
\caption{Ground state phase difference across the link $\phi_{\rm gs}$
as  a function of the ferromagnetic barrier influence $Z$,
 for $\theta=0$ and $\tilde d=0.25$.
2D $d$-wave pairing
(solid lines),
and 3D $s$-wave pairing (dashed lines).
 These  results are temperature independent.}
\end{figure}

\begin{figure}
\caption{ Ground state  phase difference $\phi_{\rm gs}$ and reduced spontaneous
current $I_{\rm gs}/I_c$ as functions of misorientation angle $\theta$
for a $d$-wave  superconducting ring interrupted by
 S/N/S,
 $Z =0$
 and S/F/S,
 $Z =3$ weak links, with  $\tilde d=0.25$ and
for three values of the reduced inverse
 inductance, $l^{-1}=0$ (open ring), 0.1 and 1.
  Low temperature, $t=0.05$ (solid curves)
   and high temperature, $t=5$ (dashed curves).}
\end{figure}

\begin{figure}
\caption{Highest spontaneous supercurrent
 $|I_{\rm gs}|/I_c$ as a  function of $l^{-1}$
in a superconducting ring interrupted by a weak link. Reduced barrier
thickness
$\tilde d=0.25$. Low temperature, $t=0.05$ (solid curves):
 (1) $Z =0$, $\theta=\pi/2$ and
 (2)  $Z =3$, $\theta=0$ for  2D $d-$wave pairing, and
 (3) $Z=3$ for  3D $s$-wave pairing. High temperature, $t=5$ (dashed curve):
 universal behavior for all above cases .}
\end{figure}

\begin{figure}
\caption{Magnetic flux $\Phi$ through a $d$-wave superconducting ring as a
function of the misorientation angle $\theta$ for $t=0.05$, $\tilde
d=0.25$, for two values of $l^{-1}$ and four values of the external
magnetic flux $\Phi_{\rm e}$.  S/N/S, $Z=0$,
and  S/F/S, $Z=3$ weak links.}
\end{figure}


\begin{thebibliography}{99}
\bibitem{zika}
We assume $\Delta_0=\Delta_0^s$ in the case of $s$-wave pairing and
$\Delta_0=\Delta_0^d$ in the case of $d$-wave pairing.
\end{thebibliography}
\end{document}